# Timing-As-A-Service (TAAS): On the role of mobile service providers in the context of integrated Industrial Internet of Things (IIoT)

Daniel Philip VENMANI[a], Fares ZERRADI[a], Fatiha HAMMA[a], Bruno JAHAN[a], Kamal SINGH[b],
*Orange, France*[a], *Telecom Saint-Etienne / University Jean Monnet*[b]

*Abstract*—Traditionally, the production efficiency of a factory floor is evaluated using non-real time objective functions. These are based on scheduling punctuality criteria such as 'earliness (a measure of finishing operations ahead of schedule)' and 'tardiness (a measure of delay in executing certain operations)'. With process automation becoming more and more inevitable due to the emergence of Industry 4.0 / Industry 5.0, real-time objective functions are also gaining popularity. One such criterion is 'synchronization' or 'timing'. In this article, an IoT enabled real-time synchronization approach is presented in the context of Integrated Industrial Internet of Things (IIoT) from a service provider's perspective. A new business and operational model, termed as "Timing-As-A-Service (TAAS)", is introduced. In this model, the role of a service provider is to "supply" synchronization from the mobile networks to the IIoT domain.

## I. INTRODUCTION

5G is the fifth generation of mobile telephony. Majority of the 5G networks worldwide are based on Time Division Duplexing (TDD). This means that every Base Station (BS), i.e., gNB in this case, should indispensably coordinate (in time and phase) with other gNBs for various reasons. These reasons can be to improve throughput, to reduce interference or to increase spectral efficiency. For such coordination to happen, every BS should be accurately 'synchronized'. Industrial Internet of Things (IIoT) [1] is one example of a use case, where the robots/machines inside a factory premises[1] require very low latency and "timing and synchronization[2]".

In this article, a new business model called *Timing-As-A-Service (TAAS)* is elaborated, which is in the context of Everything-as-a-Service (XAAS). XAAS refers to the paradigm in which various services, applications, or resources are delivered over the internet, allowing users to access and utilize them on demand [2]. At its core, this novel concept of TAAS is about real-time deployment, monitoring and control of machinery and equipment. This is done by synchronizing the robots of Non-Public Networks (NPN) at every/any stage of the production process. This in turn allows more flexibility for industrial customers in their deployment process. With the increasing emergence of use cases such as factory automation and autonomous robotics within the context of Industry 4.0 and the future Industry 5.0 [3], the idea is to have the operator serve the factory rather than the factory serving the operator. This solution is primarily meant for (small, medium size) customers who could not afford to have their own private 5G networks (i.e. principally small business owners who would not afford due to cost reasons). In such cases, the industrial customer relies on the public network operator to provide synchronization inside the factory premises. Within this context, the contributions of this paper are summarized as follows:

- A new business model is proposed based on a novel concept called 'TAAS'.
- A new gateway termed as 'smart-PTP (Smart Precision Time Protocol) gateway' is proposed to support the business model. The current state of the art from research and standards perspective lacks such device.
- The design aspects of this new device are provided in detail. The purpose of this gateway device is to create a logical over-the-air interface between the cellular system and the IIoT system.
- The system design and functional block diagram of this new smart-PTP gateway are presented.
- Results based on Matlab based simulations to illustrate the effectiveness of our proposal are provided.

The rest of the paper is organized as follows: Section I introduces the new concept TAAS. Section II dives deep into the state of the art, within the context of timing and synchronization. Sections III and IV describe the synchronization and TAAS concepts. Section V presents the smart-PTP gateway design. Section VI presents some simulated results and Section VII concludes this paper.

## II. STATE OF THE ART SYNCHRONIZATION SOLUTIONS FOR INDUSTRY 4.0 REQUIREMENTS

This section summarizes the current synchronization methods and solutions. Initially, the discussion is on the state-of-the-art synchronization solutions, which predominantly pertain to wired networks. This section then explores the progression and expansion of synchronization methods, highlighting gaps, in the context of wireless and mobile networks.

---

[1] The term 'Factory premises' is a choice made by the authors for simplicity which interchangeably refers to an Industrial campus, Non-Public Network (NPN), Private Network or Ultra-connected indoor smart factory.

[2] The terms 'Synchronization' and 'Timing' in this context are used interchangeably throughout this article with no change to their definition and meaning.

### A. Fundamentals of synchronisation

*Synchronizing in frequency:* this means that a common frequency reference (such as a common reference rhythm from Cesium/Rubidium clocks) is delivered to a set of equipment, while maintaining a given level of accuracy and stability. This type of synchronization is called "frequency synchronization", because only the frequency is matched between multiple sources. The reference signals delivered to the equipment are not necessarily aligned in phase and time.

*Synchronizing in phase/time:* this means that a common frequency reference is delivered to a set of equipment, and the reference signals are also aligned in phase and time, maintaining a given level of accuracy. An equipment which is phase-synchronized is also frequency-synchronized. The only difference consists in associating labels (e.g. timestamps) to the significant phase instants. Synchronization properties have been defined and studied by Standards Development Organizations (SDOs) like Third Generation Partnership Project (3GPP-RAN1), ITU-T Study Group (SG15) and the IEE-1588 WG. For wired networks, many protocols, e.g., IEEE 802.1 (Time Sensitive Networking (TSN)) [4], IEEE 1588 (Precision Time Protocol PTPv2) [5] have been defined and standardized. The above non-exhaustive list of protocols enables to transport and distribute synchronization through wired networks, some of which, or at times even all of them, have already been deployed in mobile operators' networks to synchronize their 4G base stations (eNBs) and/or 5G base stations (gNBs).

### B. Emerging methods of wireless synchronization

A very few solutions exist for transporting synchronization through wireless networks. To begin with, Radio Interface Based Signaling (RIBS) was studied in 3GPP RAN WG1. RIBS is a technique which enables one eNB to monitor the reference signals of another eNB by means of network listening, for performing Over-The-Air (OTA) synchronization [6]. This involves the radio interface, which is present between the eNB and the User Equipment (UE). Moving forward, there have been recent emerging proposals to adapt protocols originally developed for wired networks for use over wireless networks. In [7], the authors propose an extension to the existing IEEE 1588 PTP protocol. They propose Wireless-PTP (W-PTP) as an extension to the PTP for multi-hop wireless networks. Although there have been already some attempts in realizing PTP for wireless networks ([8], [9]), they mainly examined the limitations posed by the physical layer of wireless networks. They showed that W-PTP significantly reduces the convergence time and the number of packets required for synchronization without compromising on the synchronisation accuracy. The above-mentioned works propose various methods to transport synchronization, without any actual implementation details. In this article, a new gateway device is proposed, which is a physical hardware-based device which allows the MNOs to transport synchronization.

## III. SYNCHRONIZATION IN THE CONTEXT OF INTEGRATED 5G-IIOT APPLICATIONS

### A. Time synchronization requirements in the IIoT context

This section discusses the latest standardized accuracy values for synchronization by 3GPP. This is summarized in Table I. In this regard, 3GPP Technical Specification (TS) 22.104 [10] defines a new term called 'Clock Synchronicity'. It is defined as the maximum permissible time offset within a synchronization domain between the master clock and any individual device clock. As it could be seen in Table I, the most stringent synchronicity budget requirement is ≤ 900 ns, which corresponds to the motion-control use case.

### B. Time synchronization error budgeting in the context of Integrated 5G- IIoT Applications

Error budgeting refers to designing the end-to-end system with respect to allocating the maximum allowed time offset of the end IIoT application (as per Table I). To do this, a time synchronization model is specified in 3GPP TS 23.501 [11], as shown below.

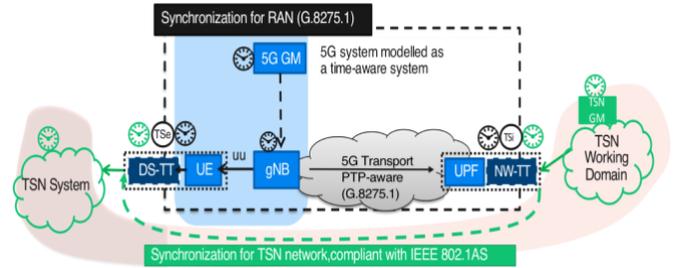

**Figure 1 : Contribution of the 5G System to the E2E sync accuracy**

From previous subsection, the maximum absolute (absolute with respect to Universal Time Coordination (UTC)) time error in order to satisfy the most stringent use case is 900 ns. In Figure 1, the goal is to be able to deliver the time error value of 900 ns between the User Equipment (UE) and the User Plane Function (UPF). This could be done, for instance, via the grand master clock (5G-GM) from the 5GS towards the TSN domain. The 5G-GM is the primary source of time for the gNBs. Generalized Precision Time Protocol (gPTP) is the protocol used for relaying master clock information from a source node to multiple end stations [11]. The connected time aware systems support gPTP. When integrating the TSN end-stations within the 5GS, it is also important to assess the impact of time error introduced between the gNB and UE over the air interface. This is illustrated in Figure 1, as 'Uu interface'.

| User-specific clock synchronicity accuracy level | Number of devices in one Communication group for clock synchronization | 5GS synchronicity budget requirement | Service area | Scenario |
|---|---|---|---|---|
| 1 (highest defined accuracy) | Up to 300 UEs | ≤ 900 ns | ≤ 100 m x 100 m | - Motion control<br>- Control-to-control communication for industrial controller |
| 2 | Up to 10 UEs | < 10 µs | ≤ 2500 m$^2$ | - High data rate video streaming |
| 3 | Up to 100 UEs | < 1 µs | ≤ 10 km$^2$ | - AVProd synchronization and packet timing |
| 4 | Up to 100 UEs | < 1 µs | < 20 km$^2$ | - Smart Grid: synchronicity between PMUs |

**TABLE I: CLOCK SYNCHRONIZATION SERVICE PERFORMANCE REQUIREMENTS FOR IIOT USE CASES (3GPP TS 23.501 [11])**

## IV. TIMING-AS-A-SERVICE (TAAS): NEW BUSINESS ENABLER

### A. Introduction to Smart PTP-Gateway (sPTP-GW)

Mobile service providers offer a range of services, including fiber connectivity, 2G/3G/4G/5G cellular connectivity, and wholesale backbone connectivity to ensure the best possible customer experience for all services. In this regard, providing 'timing' as a service to its end-customer (such as industrial players) at this point in time during the Industry 4.0 and further with future Industry 5.0 revolution could be a game changer. As defined by 3GPP in Table I, today there are several standardized industrial use cases which rely on accurate synchronization. This, in-turn, provides the opportunity for service providers to deploy cost-effective and rapidly deployable solutions to synchronize and eventually localize the machines/devices/robots inside the factory premises. To make this possible, this paper proposes a new type of equipment called '*Smart PTP-Gateway (sPTP-GW)*', as the initial step. Our proposal is based on creating a new interface. By intelligently creating an interface that integrates two different domains, i.e., the cellular domain and the IIoT domain, a new business and a new value chain for both players is created, i.e. mobile service providers as well as industrial players. Time synchronization is transported from the cellular domain to the time aware systems of the TSN domain, aka IIoT domain. A more detailed and elaborate description of this equipment is given in the next sections.

### B. Business value behind sPTP-GW

**Factory Floor:** Let us consider an example of a large car manufacturer factory with several acres of factory floor. The floor would have legacy machines/robots which are connected through a wired network. Here the connectivity between each machine is established via legacy protocols such as PROFINET. Now, the synchronization accuracy with which the legacy robots coordinate within the factory is limited to a certain accuracy value (up to 10 µs) due to the legacy wired connectivity, as well as due to the source of synchronization. First advantage of our proposal is that, through the use of the wireless connectivity provided by a device of this kind '*Smart PTP-Gateway*', the owner of the car manufacturing floor would be able to acquire the level of accuracy which is needed for a particular operation. For instance, the level of accuracy which is needed for a set of robots that are set-up for door alignment/wheel alignment would be completely different as compared to a set of cameras which are deployed in order to transmit video streams from the manufacturing floor to the control room. Second illustrative scenario would be when the car manufacturer decides to extend the factory floor by a few hundred acres. In this case, setting-up wired connectivity for a new set of robots would be a very tedious process. With a wireless gateway solution like 'sPTP-GW', factory floor owners can effortlessly scale their production floor. sPTP-GW technology makes it possible to accelerate the deployment of coordination.

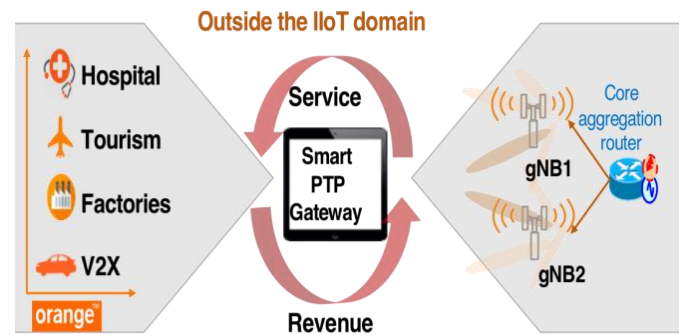

**Figure 2: Opportunities for operators in the IIoT domain**

**Airports:** As another representative example of synchronization services, airports remain a major focus for service providers. Each airport has 100s of small businesses and companies as tenants (including passengers, vehicles and IoT devices) that are coordinating with each other. All of them rely on the necessity to have an ultra-high coordination. For instance, locating a passenger or a baggage from the passenger terminal to maintenance hangars, hotels, and the car-rental facility. Achieving their smooth operation involves absolute coordination of multiple access points across multiple site locations, which in-turn necessitates an accurate synchronization among the access points. Further examples can be envisaged as shown in Figure 2.

*C. Stakeholder analysis*

The overall context is to propose and build the business value which will allow mobile operators to provide timing via mobile networks and therefore termed as "***Timing-As-A-Service (TAAS)***", a dedicated service for Industry 4.0 customers. Based on 5G/B5G networks, this new business model allows to define a high precision synchronization solution for robots automation, video streaming between control room and factory floor, positioning or indoor localization, etc. Industry sectors are often thought-of as monolithic (for instance retail, finance, oil and gas.). The reality is that each industry has multiple sub-sectors and varied types of sites, several user-groups (factory workers to control-room engineers), bunch of legacy systems, numerous applications and technology vendors and differing requirements in terms of synchronization solutions and architectures. In this context, Global System for Mobile Communications Association (GSMA) has been studying and analyzing the various examples of different 5G network options being deployed for industrial customers [12]. They have defined three different models of deployment: Public dedicated, Private and hybrid network models. Our proposed business model allows any of the players from the three groups to fully benefit without restriction, for the use cases described under Section IV. B.

*D. General system design of the sPTP-GW*

Here the basic system design of sPTP-GW is detailed. The sPTP-GW is equivalent to a 5G smart phone/tablet type of device, in the sense that its initial attach procedure follows the 3GPP 5G NR standard. Its aim is to interface the 5GS and the TSN domain. As per 3GPP TS 23.501 [11], 3GPP has defined two interfaces called 'Device-Side Time Translator (DS-TT)' and 'Network-Side Time translator (NW-TT)' at both edges of the 5GS (as seen in Figure 1). However, the DS-TT/NW-TT interfaces are designed basically to act as L2 devices, which act as a bridge between the 5GS and the TSN domain. They are designed to optionally perform link layer connectivity, discovery and reporting for discovery of Ethernet devices attached to DS-TT/ NW-TT [13]-[14].

On the other hand, the sPTP-GW is designed to act as a "smart" wireless device in order to perform time synchronization between the 5GS and the IIoT domain (some performance evaluation is provided in Section V.C). The sPTP-GW is a device (under progress in Orange), which is fundamentally a synchronization device, i.e. a device which enables to transport and distribute synchronization OTA. The distribution part itself could be based on wired or wireless networks. This device can either be a fixed device mounted across a hallway, factory floor corner or, in special cases, be a mobile device in its nature being carried by a technician across a factory floor to different locations or sites depending on the envisioned use cases. In normal operations this device will be strategically placed in a fixed location eliminating the impact of device mobility on the performance. It is evident that through this, a new "intelligent" interface is proposed between a factory floor (from the IIoT domain) and the nearest base station (of the cellular domain). It is "intelligent" in providing more flexibility, scalability and mobility to the end-stations (as illustrated in the examples above), in the sense that it is expected to serve multiple clients as well as multiple use cases with a common design-that-fit-all customers' needs. It is expected that the candidate network-based synchronization solutions should be able to meet the following conditions: (i) implementation should have low complexity to help reduce the cost, (ii) meet the current and future synchronization requirements for emerging features, (iii) provide added value with respect to existing solutions. Also all of the above should be respected without impacting either the network or the factory. Based on the criteria and the goal of avoiding the duplication of functionalities, the most viable and advantageous functionalities are tailored for future radio interface technologies to minimize costs and complexity.

Additionally, if a public network operator provides the timing service to a private factory network via the public network, the security concerns are managed by the operator. This can involve deploying a dedicated security gateway at each client's site or designating specific ports for individual factory premises, typically addressed within Service Level Agreements (SLAs).

V. TIMING-AS-A-SERVICE : TO RECEIVE AND TO RETRANSMIT

*A. Smart PTP-Gateway: Functional Block diagram*

This section proposes the functional block diagram of the PTP gateway. This is shown in Figure 3. Overall, the PTP gateway is expected to be comprised of the following functional blocks. (i) Radio frequency front and back-ends which are responsible for receiving and transmitting the synchronization radio signals; (ii) DAC and ADC converter blocks, which are responsible for converting the analog signal to digital while receiving and vice versa while transmitting. (iii) Following this is the heart of the PTP gateway, consisting in the PTP controller, the time stamping unit and the internal clock. These three functional blocks are principally responsible for time error estimation and compensation. The PTP controller is core function block which generates PTP packets. PTP stands for "Precision Timing Protocol" and is described in IEEE Standard 1588. It is a protocol for distributing time across a packet network. It works by sending a message from a master clock to a slave clock to tell the slave clock what time it is at the master. The PTP controller inside the functional block is equivalent to a PTP Telecom Boundary Clock (T-BC) for instance. A Boundary Clock (BC) is a clock node which provides mechanisms to generate PTP packets internally and then to distribute them from master port to the remaining slaves. Open source tools such as *linuxPTP* allows us to implement such a controller. This is then followed by the Time stamping unit which allows to time stamp the received *Sync and delayRequest* messages. The internal clock that gives the reference signal for the smart gateway.

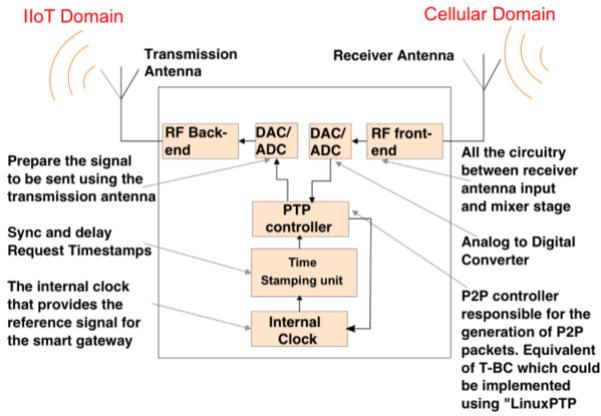

**Figure 3: Functional block diagram smart PTP gateway.**

### B. Smart PTP-Gateway: Modes of operations

Here let us dive deeper into the fundamental modes of operation of sPTP-GW. As stated previously, sPTP-GW's objective is to receive the time reference information from the nearest base station over-the-air and transport it to the IIoT domain. And therefore, two phases of synchronization are involved: To receive and to retransmit, while taking into account the total time error budget allocated for the end-to-end process. Note: As of today, this value is 900ns as noted previously. While doing so, two modes to synchronize the robots and machines inside the factory are identified: the in-band mode and out-of-band mode. These modes, illustrated in Figure 4, are described below:

*In-band mode:* The In-band mode enables delivering the timing packets on a link-by-link basis over the dedicated communication channel inside the factory site. Once sPTP-GW receives the timing packets from the nearest base station, it retransmits them over-to its closest first node inside the factory environment in order to synchronize the first machine/robot in its vicinity. This first node could be designated as the 'Master node'. This master node in-turn then retransmits the received synchronization packets towards the neighboring nodes inside the factory site. This could be done through a wired network or through wireless LAN. By doing this, it is possible to extend the distribution of synchronization to a larger network range since each node inside the factory site is responsible for receiving the timing packets from the neighboring node and then retransmitting it to its next neighbor. It is left to the factory floor owner to determine the most suitable topology within the factory for timing distribution. This mode of synchronization would provide an accurate synchronization, yet its performance depends on the number of nodes that need synchronization inside the factory site. This is because a higher number of connected devices leads to higher latency and, consequently, higher accumulated time error.

*Out-of-band mode:* The second mode is the Out-of-band mode, where the sPTP-GW is responsible for delivering the time directly to all devices within the factory site through a common communication channel. This mode eliminates the dependency on the nodes within the factory to receive and retransmit the synchronization. The sPTP-GW is solely responsible for synchronizing every node within the factory site. In this case, it requires the sPTP-GW to handle the propagation delays associated with reaching each node inside the factory since the devices may be located at different distances/zones. This results in using multiple sPTP-GW devices which are located in the vicinity of the factory nodes.

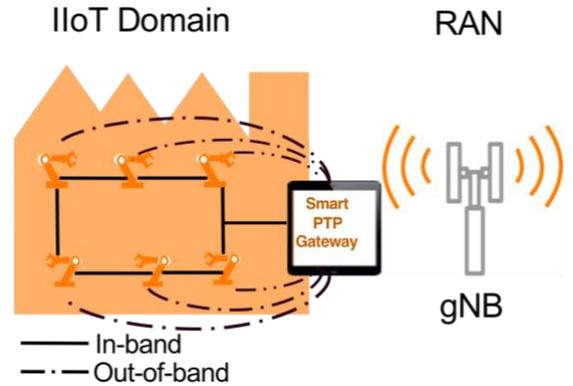

**Figure 4: Illustration of Out-band and In-band modes of operation.**

### C. Synchronization Performance

This section evaluates the efficacy of our proposal. The first step to synchronization over the radio interface is compensation of the propagation delay (PD) between gNB and sPTP-GW, by minimizing uncertainties (clock drift) between them while adhering to the existing requirements. This involves precisely studying and evaluating various PDs over radio link. PD is associated with the distance between the serving gNB and the sPTP-GW. The larger the distance, the higher the PD value.

Higher PD values result in increased time errors at robot/machine endpoints, and conversely. By carefully estimating and evaluating the PD, i.e., the time for the 5G-NR radio signal to be transmitted from the serving gNB and be received by the sPTP-GW, one could interplay with the allowed time error values (as specified in Table I) and achieve the required synchronization accuracy. This would then allow us to determine the maximum permissible distance between the gNB and sPTP-GW in order to achieve the maximum permissible time error. In order to evaluate this, the simulations were performed using MATLAB. Under the assumption of base station height of 6m and the UE height of 1.5m, the following steps were performed. *Primary Synchronization Signal (PSS)* sequence (with a length of 127) was first generated and then it was mapped to subcarriers inside the cell bandwidth. After that IFFT was applied to transform PSS sequence into signal to be transmitted. Delay, channel gain and noise were then added to the signal. Finally the signal was received and a correlation was performed with the local reference sequence. This was used for peak detection and finally the peak delay was retained. The simulations used the TDL-C pathloss model summarized in 3GPP TR 38.901, with a delay spread of 300 ns [15].

Based on our simulations, the error budget for transporting the timing packets over radio interface is analyzed by detecting the PSS sequence of the 5G-NR signal. In 5G NR system, PSS and Secondary Synchronization Signal (SSS) are used to detect the best serving gNB. Evaluations for Sub-Carrier Spacing (SCS) of 15 kHz, 30 kHz and 60 kHz for 6 GHz frequency and FFT = 4096 are presented here. We show the worst timing error achieved for 90[th] percentile cases for different SCS and range (maximum distance between gNB and sPTP-GW). The 90[th] percentile corresponds to 5000 simulation runs. Our results in Table II show that, for a given error budget, the mean error will be determined by the ideal choice of SCS and maximal distance between gNB and sPTP-GW. The timing error increases with the distance between gNB and sPTP-GW due to multipaths and is also dependent on the SCS values. It can be seen that the time error is similar for different SCS values when the range is 1 Km. Though it is slightly lower for higher SCS values in this case. However, this changes for higher ranges such as 3 Km where lower SCS values show better robustness against synchronization failures and this also relatively lowers their time error. This is because with increasing SCS, the cyclic prefix interval decreases which leads to increased Inter Symbol Interference during poor radio conditions. This results in synchronization failure when looking at the worst 90[th] percentile performance with SCS value of 60 kHz. Thus, a given error budget will determine the ideal choice of SCS for a given maximal distance between gNB and sPTP-GW. It can also be noted that the timing error is largely inferior to 900ns, for SCS values of 15kHz and 30 KHz. Hence, these SCS values satisfy the synchronization requirements specified in Table I.

Note that in our simulations the delay estimation is based on peak detection after correlation. In practice, advanced algorithms can be designed to detect the first path even if it does not correspond to the peak. This will reduce the time to detect the peak and therefore enhance the delay estimation accuracy. Moreover, more appropriate channel models (which suit best for each individual Industry scenario) can be designed for evaluating time synchronization mechanisms. This will bring the simulations closer to reality.

| Sub-Carrier Spacing | 90[th] percentile worst timing error with 1km range | 90[th] percentile worst timing error with 3km range |
|---|---|---|
| 15 kHz | 30.4 ns | 244.1 ns |
| 30 kHz | 26.7 ns | 391.8 ns |
| 60 kHz | 23.6 ns | Synchronisation failure |

**Table II: Timing error for different Sub-Carrier Spacing and range (maximum distance between gNB and sPTP-GW)**

*D. Research Challenges and Future Directions*

The innovation behind the idea to synchronize the machines/robots inside a factory using directly the gNBs of a public network in vicinity is presented here. Some open research challenges in that direction are discussed below:

- One of the causes of time error is multipaths in wireless environments which arrive with different delay spreads. One open issue is to design efficient Time of Arrival (ToA) detection methods which can minimize time error by detecting the first PSS signal using the first "fastest" path in case of multipaths.
- There exist different channel models such as described in 3GPP TR 38.901 [15]. Nevertheless, as margins of the order of nanoseconds are targeted, another open issue is to design new advanced and precise models such as ray tracing models considering Refractive Index Surface (RIS) that are suited to time synchronization.
- Our case here includes one gNB sending-out PSS signals towards the factory. One research issue here is to study if multiple gNB clusters could coordinate in order to improve the time synchronization for the robots of the factory.
- Furthermore, other factors such as interference and mobility within factory and production floors impact accuracy and clock skew. Wireless signals are susceptible to interference especially in a volatile and chaotic factory floor and therefore offer poor timing coordination among large, interconnected devices in mission critical applications. Time synchronization should be designed in a way that it is efficient even in case of interference and more specialised channel models are needed to simulate electromagnetic interference found in industrial environments. Also, sPTP-GW typically remains stationary during normal operations. This eliminates any potential impact of mobility on the performance within the cellular domain. Robust mechanisms to counter the impact of mobility will have to be designed for the specific case where sPTP-GW needs to be mobile.

## VII. Conclusion

A successful telecoms operator will profitably contribute to improving society by enabling enterprises, consumers and industries to collaborate in a more efficient way. In this context, the concept of TAAS is presented along with a new business model. This business model is complemented through the conception of a smart device which will allow industrial machines and robots to be connected and communicate with the mobile operator network. The effectiveness of this solution is evaluated using simulations. Results demonstrate that the 90th percentile timing error (using SCS of 15 kHz and 30 kHz) falls well below the stringent timing synchronization threshold of 900 ns, essential for Industrial Internet of Things (IIoT) use cases. This holds true even when the distance separating the gNB and the proposed sPTP-GW reaches a substantial 3 Km. Consequently, this solution proves highly advantageous for factories situated in remote, rural environments, ensuring reliable performance even in such challenging geographical locations.

## REFERENCES

[1] Mourtzis D, Angelopoulos J, Panopoulos N. Smart Manufacturing and Tactile Internet Based on 5G in Industry 4.0: Challenges, Applications and New Trends. Electronics. 2021; 10(24):3175.
[2] Chang, Z., Zhou, Z., Zhou, S., Chen, T., & Ristaniemi, T. (2017). Towards service-oriented 5G: Virtualizing the networks for everything-as-a-service. IEEE Access, 6, 1480-1489.
[3] Leng, J., Sha, W., Wang, B., Zheng, P., Zhuang, C., Liu, Q., Wuest, T., Mourtzis, D. and Wang, L., 2022. Industry 5.0: Prospect and retrospect. Journal of Manufacturing Systems, 65, pp.279-295.
[4] Nasrallah, Ahmed, et al. "Ultra-low latency (ULL) networks: The IEEE TSN and IETF DetNet standards and related 5G ULL research." IEEE Communications Surveys & Tutorials 21.1, 88-145, 2018.
[5] J.C. Eidson, M. Fischer, and J.White. "IEEE-1588™ Standard for a precision clock synchronization protocol for networked measurement and control systems." Proceedings of the 34th Annual Precise Time and Time Interval Systems and Applications Meeting. 2002.
[6] Mahmood, Aamir, et al. "Over-the-air time synchronization for URLLC: requirements, challenges and possible enablers." 2018 15th International Symposium on Wireless Communication Systems (ISWCS), 2018.
[7] A. Garg, A. Yadav, A. Sikora and A. S. Sairam, "Wireless Precision Time Protocol," in IEEE Communications Letters, vol. 22, no. 4, pp. 812-815, April 2018.
[8] H. Cho, J. Jung, B. Cho, Y. Jin, S.-W. Lee and Y. Baek, "Precision time synchronization using IEEE 1588 for wireless sensor networks", Proc. Int. Conf. Comput. Sci. Eng., vol. 2, pp. 579-586, Aug. 2009.
[9] W. Wallner, et al. "A simulation framework for IEEE 1588", Proc. IEEE Int. Symp. Precis. Clock Synchronization Meas. Control Commun. (ISPCS), pp. 1-6, Sep. 2016.
[10] 3GPP TS 22.104, Service requirements for cyber-physical control applications in vertical domains, V19.0.0, Mar. 2023.
[11] 3GPP TS 23.501, System architecture for the 5G System (5GS), V18.2.1, June 2023.
[12] *"Mobile networks for Industry 4.0"*. https://www.gsma.com/iot/networks-industry40/ (accessed June 29, 2023).
[13] L. Grosjean, et al. 5G-Enabled Smart Manufacturing-A booklet by 5G-SMART. arXiv preprint arXiv:2209.10300, 2022.
[14] Noor-A-Rahim, Md, et al. "Wireless Communications for Smart Manufacturing and Industrial IoT: Existing Technologies, 5G and Beyond." Sensors 23.1 pp.73, 2022.
[15] 3GPP TR. 38.901, Study on channel model for frequencies from 0.5 to 100 GHz, V17.0.0, Mar. 2022.

BIOGRAPHY

Daniel Philip Venmani is a research and standardization engineer at Orange. His primary research activities focus on the development of future mobile transport network architectures. He is very actively involved in standardization activities – representing Orange at ITU-T Study Group 15, 3GPP SA2, IETF, IEEE1588, IEEE802.1CM, and more. He obtained his Ph.D. from UPMC-Paris VI, France.

Fares Zerradi is currently working as radio engineer with Orange. He was previously a researcher in timing and synchronization aspects, developing novel algorithms for 5G timing system aspects. He obtained his master's degree from the University of Montpellier.

Fatiha Hamma received an M.S degree in embedded and mobile system engineering from the National School of Computer Science and Systems Analysis, Rabat, Morocco, and a master's degree in Engineering in Computer Science from Le Mans Université, Le Mans. In France, she is currently a Ph.D. student at University Jean Monnet with Orange, Lannion.

Bruno Jahan received an M.S. degree in optical and photonics and another M.S. degree in electronic systems from the University of Paris-Saclay, France, in 1989 and 1990, respectively. In 1991, he was with Telediffusion de France as a Research Engineer. He joined Orange Labs (formerly France Telecom), Rennes, in 1998. His research interests include digital signals processing for wire and wireless communications.

Kamal Singh received the B.Tech. degree in electrical engineering from IIT Delhi, India, in 2002, and the Ph.D. degree in computer science from the University of Rennes 1, Rennes, France, in 2007. He then worked as a postdoctoral researcher at INRIA and IMT Atlantique, Rennes. He is currently an Associate Professor with the Université de Saint-Étienne, France, working in the team Data Intelligence of Laboratoire Hubert Curien. His research interests include the Internet of Things, mobile networks and AI.